\documentclass[12pt,preprint]{aastex} 
 
\title{Hot Horizontal Branch stars in $\omega$ Centauri: clues about their origin 
from the cluster Color Magnitude Diagram}

\author{Santi Cassisi}
\affil{Osservatorio Astronomico di Teramo, Via M. Maggini, 64100
Teramo, Italy; cassisi@oa-teramo.inaf.it}

\and

\author{Maurizio Salaris}
\affil{Astrophysics Research Institute, Liverpool John Moores
University, Twelve Quays House, Birkenhead, CH41 1LD, UK;
ms@astro.livjm.ac.uk}

\and

\author{Jay Anderson}
\affil{Space Telescope Science Institute, 3800 San Martin Drive,
Baltimore, MD 21218, USA; jayander@stsci.edu}

\and

\author{Giampaolo Piotto}
\affil{Dipartimento di Astronomia, Universit\'a di Padova, Vicolo
dell'Osservatorio 2, I-35122 Padova, Italy; giampaolo.piotto@unipd.it}

\and

\author{Adriano Pietrinferni}
\affil{Osservatorio Astronomico di Teramo, Via M. Maggini, 64100
Teramo, Italy; pietrinferni@oa-teramo.inaf.it}

\and

\author{Antonino Milone}
\affil{Dipartimento di Astronomia, Universit\'a di Padova, Vicolo
dell'Osservatorio 2, I-35122 Padova, Italy; antonino.milone@unipd.it}

\and

\author{Andrea Bellini} 
\affil{Space Telescope Science Institute, 3800 San Martin Drive,
Baltimore, MD 21218, USA; bellini@stsci.edu}
 
\and

\author{Luigi R.\ Bedin} 
\affil{Space Telescope Science Institute, 3800 San Martin Drive,
Baltimore, MD 21218, USA; bedin@stsci.edu}

\begin{document}

\begin{abstract}
 
\noindent

We investigate a peculiar feature at the hottest, blue end of the horizontal branch of 
Galactic globular cluster $\omega$ Centauri, using the high-precision and nearly complete 
catalog that has been constructed from a survey taken with the {\sl Advanced Camera for Survey} 
on board the Hubble Space Telescope, that covers the inner 
$10\times 10 $ arcminutes.
It is a densely
populated clump of stars with an almost vertical structure in the
F435W-(F435W$-$F625W) plane, that we termed \lq{blue clump}\rq.  A
comparison with theoretical models leads to the conclusion that this
feature must necessarily harbor either hot flasher stars, or canonical
He-rich stars --progeny of the blue Main Sequence sub-population
observed in this cluster-- or a mixture of both types, plus possibly a
component from the normal-He population hosted by the cluster.  A
strong constraint coming from theory is that the mass of the objects
in the \lq{blue clump}\rq\ has to be very finely tuned, with a spread
of at most only $\sim$0.03~$M_{\odot}$.  By comparing observed and
theoretical star counts along both the H- and He-burning stages we
then find that at least 15\% of the expected He-rich Horizontal Branch
stars are missing from the color-magnitude diagram.  This missing
population could be the progeny of red giants that failed to ignite
central He-burning and have produced He-core White Dwarfs. Our
conclusion supports the scenario recently suggested by Calamida et
al.\ (2008) for explaining the observed ratio of White Dwarfs to Main
Sequence stars in $\omega$ Centauri.
\end{abstract}
 

\section{Introduction}

\noindent
The origin and evolution of the Galactic Globular Cluster (GGC)
$\omega$ Centauri (NGC 5139) is one of the most exciting open problems
of modern astrophysics.  $\omega$ Centauri belongs to the Galactic GC
system but, due to its complex stellar population and unusual large
mass ($M\sim3\times10^6M_\odot$) it has been often considered the
relic of a larger stellar system, such as a dwarf galaxy (see Lee et al.\ 1999,
Villanova et al.\ 2007 and references therein).  In these last few
years, the availability of high- accuracy photometry and multi-object
spectroscopy prompted the discovery of additional unusual features
like the discrete structure of its red giant branch (RGB -- Lee et
al.\ 1999, Pancino et al.\ 2000, Rey et al.\ 2004, Sollima et al.\
2005), the peculiar -- and still not poorly understood -- age
distribution (Sollima et al.\ 2007, Villanova et al.\ 2007 and
references therein), the distinct kinematical properties of the
various stellar components (Sollima et al.\ 2005).  An additional
level of complexity has been added by the discovery by Anderson (1997)
and Bedin et al.\ (2004) that, over a range of at least 2 mag, the
cluster main sequence (MS) splits into two separate sequences, the
bluer MS being more metal-rich than the red one (Piotto et al.\ 2005).
Apparently, the only way to account for both the spectroscopic and
photometric observations is to assume that the blue MS is populated by
stars with an extremely large initial helium content ($Y\sim0.38$
--Piotto et al.\ 2005, Lee et al.\ 2005).

$\omega$ Cen also displays an \lq{intriguing}\rq\ horizontal branch
(HB), characterized by a very extended blue tail, with gaps (see
Whitney et al.\ 1994 and references therein).  In addition, $\omega$
Cen, together with NGC 2808 (Brown et al.\ 2001), NGC 6715 (Rosenberg, Recio-Blanco \& Garc{\'i}a-Mar{\'i}n\ 2004), NGC 6388 and possibly NGC 6441 (Busso et al.\ 2007),  shows
\lq{blue hook}\rq\ stars at the very hot end of its HB (Whitney et
al.\ 1998, D'Cruz et al.\ 2000). This feature consists of a
significant number of objects located 
at hotter effective temperatures than the hot end of the HB
observed in other GCs with blue tails, such as NGC 6752 (Momany et al.\ 2002, 2004) and fainter 
than the bluest end of the canonical zero age
HB (ZAHB), by up to about 0.7 mag in the UV photometric filters.  The
presence of these peculiar stellar objects cannot be explained by
canonical stellar evolution (Salaris \& Cassisi 2005).

A few years after their discovery, Brown et al.\ (2001) proposed to explain the 
origin of \lq{blue hook}\rq\ stars via the \rq{He-flash induced mixing}\rq\ scenario. 
This
scenario was first envisaged by Castellani \& Castellani~(1993), who proposed that 
stars that lose an
unusually large amount of mass along the RGB will leave the branch
before the occurrence of the He flash, and will move quickly to the
(helium-core) white dwarf cooling curve before eventually initiating
helium burning.  The evolution of these so-called \lq{late hot helium
flashers}\rq\ differs significantly from that of stars which undergo
the helium flash at the tip of the RGB. In fact, in this case the large entropy 
barrier of the efficient H-burning shell prevents the products of helium fusion from being
mixed within the surface convective region. These stars will thus
evolve to the ZAHB without any change in their envelope
composition. By contrast, stars that ignite helium on the white dwarf
(WD) cooling curve have a much less efficient H-burning shell, and
will undergo extensive mixing between the He-core and the H-rich
envelope, as predicted by Sweigart (1997) and Brown et al.\ (2001).
The exact development of the He-flash induced mixing has been
investigated on the basis of fully evolutionary models firstly by
Cassisi et al.\ (2003) and more recently by Miller Bertolami et al.\
(2008). The exact chemical composition of the envelope at the end of
this process depends on the strength of the He-flash, i.e. on when --
along the WD cooling sequence --the He ignition occurs. 
Irrespective of
the computational details, all investigations predict a large increase
of the surface abundance of He and a significant enhancement of the
carbon abundance in the envelope.

An alternative scenario for explaining the presence of \lq{blue
hook}\rq\ stars is related to the observed split of the MS. Lee et
al.\ (2005) were the first to suggest that \lq{blue hook}\rq\ stars
are the progeny of the He-rich MS. At odds with the He-flash induced
mixing scenario, if \lq{blue hook}\rq\ stars were to be explained by
the helium-enrichment scenario, their surface He abundance should not
exceed the value estimated for the blue MS ($Y\approx$0.38) and C
should not be enriched at all.  Interestingly enough, we now know that
NGC 2808, the other GC hosting \lq{blue hook}\rq\ stars, shows also
the splitting of its MS in three distinct sequences (D'Antona et al.\
2005, Piotto et al.\ 2007).

Early spectroscopic measurements for 12 \lq{blue hook}\rq\ candidates
performed by Moehler et al.\ (2002) confirmed that these stars are
indeed both hotter and on average more He-rich than the canonical
extreme HB stars. A more recent analysis (Moehler et al.\ 2007) has
shown that at least $\sim$15\% of the hottest HB stars in $\omega$ Cen
have a surface He abundance larger than predicted by the He enrichment
scenario. In addition, the most He-rich stars also have enhanced
surface carbon abundance, in good agreement with the He-flash induced
mixing scenario. From these spectroscopic observations, it appears that
the He enrichment scenario cannot account for all \lq{blue hook}\rq\
stars.

Quite recently, an additional puzzling result has been obtained by
Calamida et al.\ (2008).  They measured the ratio of White Dwarfs (WDs)
to MS star counts to be is at least a factor of two larger than the
theoretical value predicted from the ratio of Carbon-Oxygen (CO) core
WD cooling times to MS lifetimes.  This discrepancy can be solved only
by assuming the presence of a significant fraction --ranging from 10\%
to 80\%, depending on their masses-- of He-core WDs.

To gain additional information about the evolutionary properties of
the various subpopulations harbored by this cluster, we take advantage
of a rich and nearly complete ACS on board HST photometric optical data-set for the HB
of $\omega$ Cen. The color-magnitude-diagram (CMD) shows a prominent
and well populated feature at the faint end of the HB. The density of
stars in this feature is definitely higher than immediately beyond its
brighter, cooler end.  Our aim is to investigate how this feature
relates to the \lq{blue hook}\rq\ discovered in UV filters and, 
more generally,
what are the constraints imposed by the observed HB population on the
progeny of the blue MS subpopulation.  Section~2 describes the data
and an analysis of the \lq{blue hook}\rq\ population based on the
comparison with suitable evolutionary models. In Section~3, we compare
observed star counts along the HB with theoretical expectations for
the ratio between the progeny of the blue and red main sequences; a
summary follows in Section~4.

\section{The photometric data and comparison with stellar models}
 
The data set used in this work consists of a mosaic of $3\times3$
fields obtained with the ACS/WFC (GO-9442, PI A.\ Cool) through the
F435W and F625W filters.
Each of these fields has one short (about 10 s) and three long
exposures (about 340 s) in both F435W and F625W.  The ACS/WFC mosaic
covers the inner $\sim 10\arcmin \times 10\arcmin$ centered on the
cluster.
The images were reduced and presented in Anderson \& van der Marel
(2009) using the software described in great detail in Anderson et
al.\ (2008).
Briefly, the program finds and measures each star in all exposures
simultaneusly by fitting a spatially-variable effective point-spread
function.
Instrumental magnitudes were transformed into the ACS Vega-mag flight
system following the procedure given in Bedin et al.\ (2005), using
the zero points of Sirianni et al.\ (2005).  

The PSFs used here were constructed from the bright, isolated stars in the many exposures, using the approach discusses in Anderson \&  King (2006), wherein we model PSF as the sum of a spatially  variable but temporally constant part and a spatially constant but temporally variable perturbation, to account for breathing-related focus changes.  Even this sophisticated model cannot account for  all the subtle variations of the HST PSF, so to deal with the residual PSF errors and to account for possible reddening variations  across the field, we used a method similar to that 
adopted in Milone et al. (2008) and Sarajedini et al (2007). This method involves fitting the average cluster sequence with a MSRL (main-sequence ridgeline), examining how the stars within a local region may lie systematically to the blue or red of this sequence, and constructing a spatial correction to remove this systematic trend. The maximum correction made here was 0.034 mag in color, but is typically less than 0.01 mag.  Artificial-star tests indicate that the completeness is greater than 95\% for all stars along the HB.

\begin{figure}        
\plotone{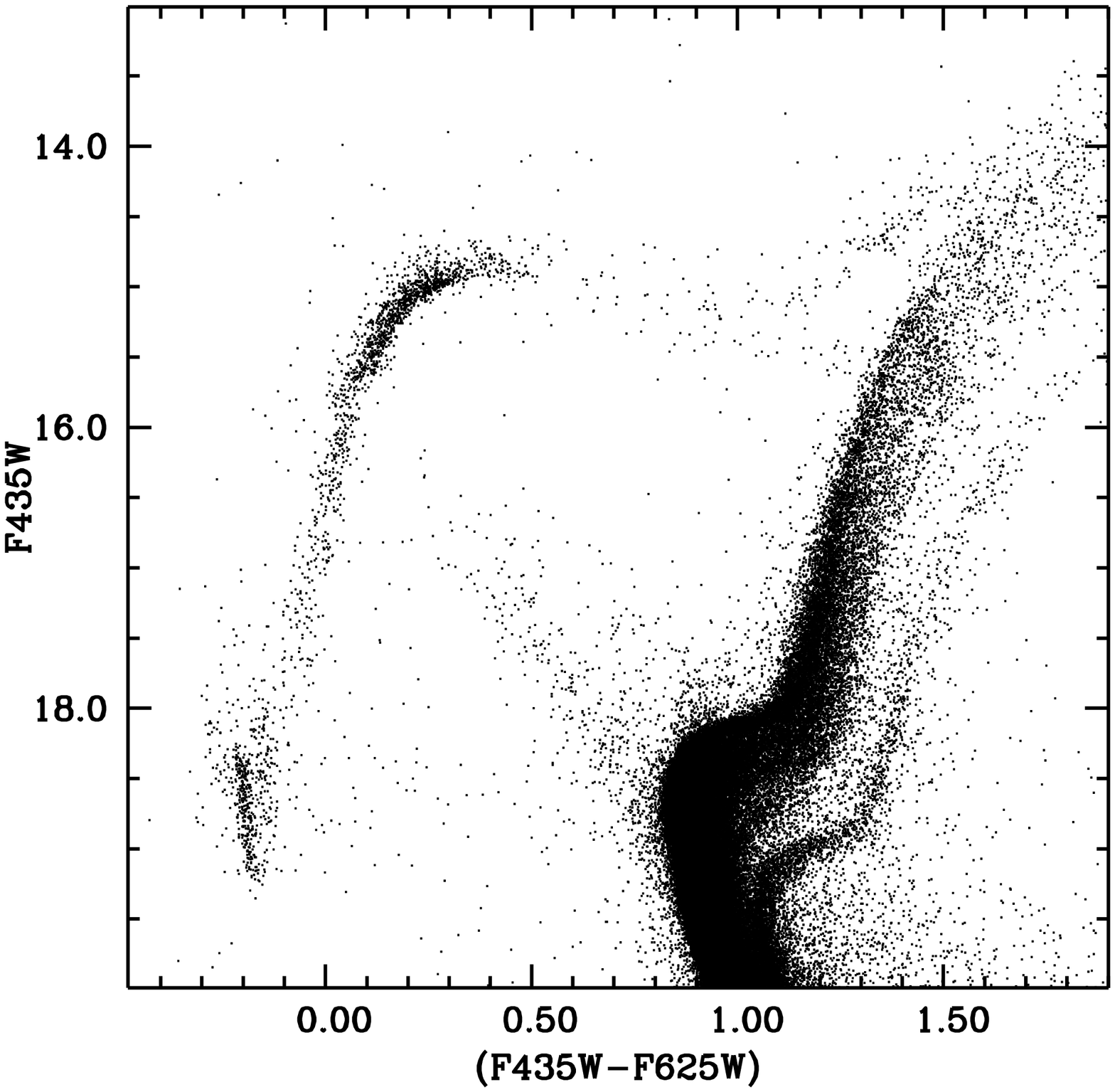}        
\caption
{The Color-Magnitude diagram of $\omega$~Cen used in our analysis.
\label{cmd1}}        
\end{figure}

The CMD is displayed in Fig.~\ref{cmd1} and shows the well known,
complex stellar population harbored by the cluster
(see, e.g. Villanova  et al.\ 2007 for a brief summary).  
Here we focus our attention on the HB, populated by about 2200 objects
in our CMD.  To start our analysis, we derive a distance modulus and
reddening by fitting ZAHB models from the BaSTI database by
Pietrinferni et al.~(2006) for [Fe/H]=$-$1.62 ([$\alpha$/Fe]=0.4) and
initial He mass fraction $Y$=0.246 (hereinafter they will be denoted as
'reference' models) to the observed HB.  This chemical composition
is consistent with the metal content of the main stellar component of
$\omega$ Centauri ([Fe/H]$\sim -$1.7, e.g.\ Villanova et al.~2007,
Sollima et al.~2008).  We derive $(m-M)_{\rm F435W}$=14.32 and
E(F435W$-$F625W)=0.22. Following the results by Bedin et al.~(2005),
this reddening corresponds to E(B-V)=0.15 and $A_{\rm F435W}$=0.63
using the Cardelli, Clayton \& Mathis (1989) extinction law with
$R_V$=3.1. The estimated true distance modulus is therefore
$(m-M)_{0}$=13.69.  This quick estimate of distance and reddening is
sufficient for the purposes of this investigation, and compares very
well with $(m-M)_{0}=13.66\pm0.12$ obtained from the work by Thompson
et al.~(2001) on the cluster eclipsing binary OGLEGC17, using surface
brightness vs.\ infrared color relationships, and their assumed
reddening E(B-V)=0.13$\pm$0.02. It is also consistent with the
estimate $(m-M)_0=13.70\pm0.06\pm0.06$ (random and systematic error)
obtained by Del Principe et al.\ (2006) employing a near-infrared
Period-Luminosity relation of RR Lyrae stars.

We wish to notice here that the models employed in this
distance/reddening estimate and in the rest of this paper do not take
into account the effect of radiative levitation of metals in the
atmospheres of HB stars, that appears to be efficient in objects with
$T_{\rm eff}$ hotter than $\sim$11$\,$000~K (see, e.g. Gr\"undahl
et al.\ 1999, Behr~2003 and references therein). To have a first order estimate
of the magnitude of this effect, we have transformed the theoretical
ZAHB models with $T_{\rm eff}$ larger than 11$\,$000~K (corresponding 
approximately to the ZAHB location of the 0.6$M_{\odot}$ model)
to the photometric bands of our CMD, using bolometric corrections
computed for [Fe/H]=0.0, i.e. an iron abundance consistent with 
the spectroscopic measurements by Behr~(2003) in globular cluster HB 
stars hotter than $\sim$11$\,$000~K. 
This fictitious ZAHB with enhanced metals
in the atmosphere overlaps almost perfectly with the canonical one in
the relevant $T_{\rm eff}$ range, differences in the F435W
magnitudes [at fixed (F435W$-$F625W) color] being of the order of
0.05~mag (enhanced metal models being brighter) decreasing when
moving to the blue end of the HB.

Figure~\ref{cmd2} displays the fit of our reference ZAHB (solid line)
to the photometry of the HB stars. The constraint on $(m-M)_{\rm
F435W}$ comes essentially from the overdensity of stars at and around
the 'knee' of the observed HB distribution [i.e., around (F435W$-$F625W)$\sim 0.2$], 
whereas the reddening is constrained by the fit to the bright vertical
part of the blue HB.  The stars at the red end
of the 'knee' [F435W$-$F625W) between $\sim$0.3 and 0.5 mag] 
appear to be largely evolved off the ZAHB, and the red part of the HB
is only very sparsely populated. Beyond the blue edge of the 'knee'
the number of HB stars also decreases sharply.  Three selected HB
evolutionary tracks for 0.57, 0.60 and 0.63 $M_{\odot}$ are also
plotted, to show the approximate mass range corresponding to the
'knee' of the HB distribution.  At the blue end of the observed HB we
have marked (thin solid line) the boundary of a well populated (about
400 objects) almost vertical, clump-like feature, that extends also well
below the theoretical ZAHB, by $\approx$0.5~mag in F435W. The
density of stars in the feature is definitely larger than on the
cooler side of the indicated redder limit.  The $T_{\rm eff}$ of the
objects populating this feature is larger than $\approx$30$\,$000~K,
using the colour-$T_{\rm eff}$ relation from the reference ZAHB.  No
reasonable variations of distance modulus/reddening/metallicity can
change the fact that this feature extends to magnitudes fainter than
ZAHB models. This occurrence mirrors the discovery of \lq{blue
hook}\rq\ stars in far-UV observations of this cluster (Whitney et
al.\ 1998; D'Cruz et al.\ 2000) i.e.\ a population of hot HB stars,
lying beyond canonical ZAHB models.

The large population of HB stars in our CMD allow us to investigate the origin of the 
objects populating this feature. From now
on we will denote them as \lq{blue clump}\rq\ stars.  The left panel
of Fig.~\ref{cmd3} displays an enlargement of the \lq{blue clump}\rq\
region, with overimposed the ZAHB and evolutionary track of the least
massive ($M=0.491~M_{\odot}$) canonical HB model (solid line) for the
reference composition [Fe/H]=$-$1.62 ([$\alpha$/Fe]=0.4, Y=0.246).  The plot
also shows a hot flasher model (dotted line) from Cassisi et
al.~(2003) for a mass typical  ($M=0.489~M_{\odot}$) of the hot flasher progeny of
the canonical, reference ([Fe/H]$\sim-$1.7) population. 
Two $\alpha$-enhanced HB models (dashed lines) plus the ZAHB locus  
representative of the He-rich subpopulation with $Z$=0.0016 and $Y$=0.4 (corresponding to [Fe/H]$\sim -1.3$).
These He-enhanced HB models would correspond to the progeny of the blue MS stars
identified by Bedin et al.\ (2004) therefore their iron content has been chosen consistent
with the [Fe/H] estimate obtained for the blue MS stars by Piotto et al.\ (2005).

Employing only reference HB models does not allow to reproduce this
feature.  Photometric errors at these magnitudes are equal to only
$\sim 0.01$~mag in both filters, therefore the existence of largely
underluminous stars ($\approx0.5$ mag) below the canonical reference
ZAHB cannot be explained in terms of a photometric spread.  The
hot-flasher model, with $M=0.489~M_{\odot}$, can reach fainter F435W
magnitudes, and is therefore a better candidate to explain the
presence of these stars. We also wish to notice that the use of models
with different total mass for the hot flashers would not modify
the outlined scenario since, as has been shown by Brown et al.\
(2001, but see also Miller Bertolami et al.\ 2008) all hot
flasher models tend to cluster in a
very narrow region of the CMD, regardless of their total mass. 
 
Canonical HB models originated from a He-rich population can also
reproduce the bulk of the \lq{blue clump}\rq\ stars (although, as for
the case of hot flashers, a few objects remain underluminous compared
to theoretical models).  The two models displayed in Fig.~\ref{cmd3}
have masses equal to 0.460 and 0.461 $M_{\odot}$, respectively.

As a note of caution we recall here that all the evolutionary
sequences have been translated from the theoretical H-R diagram to the
observational CMD by using the color-$T_{\rm eff}$ relation and
bolometric corrections provided by Bedin et al.\ (2005), based on
synthetic spectra computed by assuming scaled solar abundances.  As
shown by Brown et al.\ (2001), these spectra do not represent
accurately the peculiar atmospheres of hot flashers, because these
stars are expected to have strongly enhanced He and C abundances at
the surface. More in detail, for the atmospheres of hot flashers,
Brown et al.\ (2001 -- see their Fig.~11) have shown that an
atmospheric composition of 96\% He and 4\% C (or N) produces lower
fluxes in the F435W filter, compared to a standard metal mixture.
Therefore, for the hot flasher model one should use more appropriate
--but not yet available-- bolometric corrections, that however we do
not expect to alter the basic conclusions of this analysis.  As for
the He-rich stellar population, Girardi et al.\ (2007) have shown that
the effect of an enhanced He content --of the order of $\Delta
Y\sim0.1-$0.2-- on bolometric corrections and colors is negligible at
the $T_{\rm eff}$ values corresponding to hot HB stars.

To summarize the result of the previous comparisons, the \lq{blue
clump}\rq\ feature in our optical CMD appear to be reproduced by
either hot flashers or canonical He-rich HB stars or a mixture of both
types, plus possibly also a contribution, at least for the brighter
part, from the reference HB population.  {\it The striking result is
that such a sharp feature can be produced only by stars in an
extremely narrow range of masses, up to at most only 0.03$M_{\odot}$,
according to the models discussed above.}

To corroborate these conclusions, based only on the comparison of
individual HB tracks and ZAHB models, we have computed some synthetic
HB models taking into account the \lq{spectroscopical
classification}\rq\ of the hottest stars along the HB of $\omega$ Cen
obtained by Moehler et al.~(2007). They found that out of $\sim$40
objects hotter than 30$\,$000~K, at least $\sim$15\% show extreme
enhancements of He and should be the observational counterpart of hot
flashers, $\sim$35\% are He-poor stars that we associate to the
reference HB models, and $\sim$50\% have enhanced He but below the
value for the hot flashers, and we identify them as canonical HB stars
produced by the He-rich population. Some of the objects in this last
group could also be hot flashers, since they show He-abundances above
$Y$=0.40.

The observations by Moehler et al.~(2007) refer to a different cluster
field and may be affected by completeness effect, but we use them here
just to show how a mixture of these three types of objects can
qualitatively reproduce this CMD \lq{blue clump}\rq\ . We calculated a
synthetic CMD using the same methods described in
Salaris, Cassisi \& Pietrinferni~(2008),  
and employing the population ratios among the various components as given
above.  For the He-rich population we have considered the
0.460$M_{\odot}$ model, and for the reference HB population the
0.491~$M_{\odot}$ model. We have included photometric errors coming
from the data reduction, that are of the order of 0.01~mag.  The right
panel of Fig.~\ref{cmd3} displays the synthetic CMD, with the same
number of stars as observed, that shows clearly a CMD feature very
similar to the observed \lq{blue clump}\rq\ .

\section{Star counts along the HB}

As mentioned in the previous section, the range of masses involved populating the \lq{blue
clump}\rq\ in our optical CMD must be extremely narrow, $\sim$ 0.03
~$M_{\odot}$ at most.  Star counts along the observed HB displayed in
Fig.~2 provide a fraction of \lq{blue clump}\rq\ stars to redder HB
objects equal to 22$\pm$2 \%. The error estimate includes in
quadrature the Poissonian error on the star counts and the error
deriving from an uncertainty of $\pm$0.1~mag (in F435W) on the
placement of the boundary of this feature.  

As discussed before, the \lq{blue clump}\rq\ in our CMD may be partly made of the progeny of the cluster He-rich
population. It is therefore necessary to investigate whether the
fraction of HB stars in the \lq{blue clump}\rq\ is compatible with the
fraction of He-rich stars seen along the MS phase.  To answer this
question we have estimated theoretically the fraction of HB stars
expected to be generated by the He-rich subpopulation. We have first
calculated the differential luminosity function (i.e. number of stars
per magnitude interval) for both an isochrone with the reference
composition and a He-rich one (same compositin of the He-rich HB models) 
from the MS to the tip of the RGB.  We have used a Salpeter mass
function and assumed that the two populations are coeval, with an age
of 13.5~Gyr.  The relative number of MS stars in the two populations
has been normalized by requiring that the number ratio of the He-rich
to the reference MS population is equal to 0.92, consistent with the
value observed in this field by Bellini et al.\ (2009) 
for the stars in the magnitude interval between F435W=20.9
and 22.1.  With the normalization in place, the theoretical luminosity
functions provide the ratio of objects at the tip of the RGB belonging
to the two populations. This value corresponds, in turn, to the ratio
of the rates with which RGB stars of both populations are fed onto the
HB.

It turns out that the rate for the He-rich component is about 30\% of
the value for the reference population. This result is practically
unaffected when we decrease the age by $\sim$1~Gyr for the He-enhanced
population. An age decrease of this order of magnitude is expected in
case the latter population is produced from the ejecta of stars
belonging to the reference population. The result is also only
marginally affected by the choice of an initial mass function
different than the Salpeter's one, because the mass range spanned in
the MS normalization is not large. We have recomputed this ratio
using a flat mass function and one with a power-law exponent $-$3. The
resulting ratio changes by only $\pm$0.5\%.

\begin{figure}        
\plotone{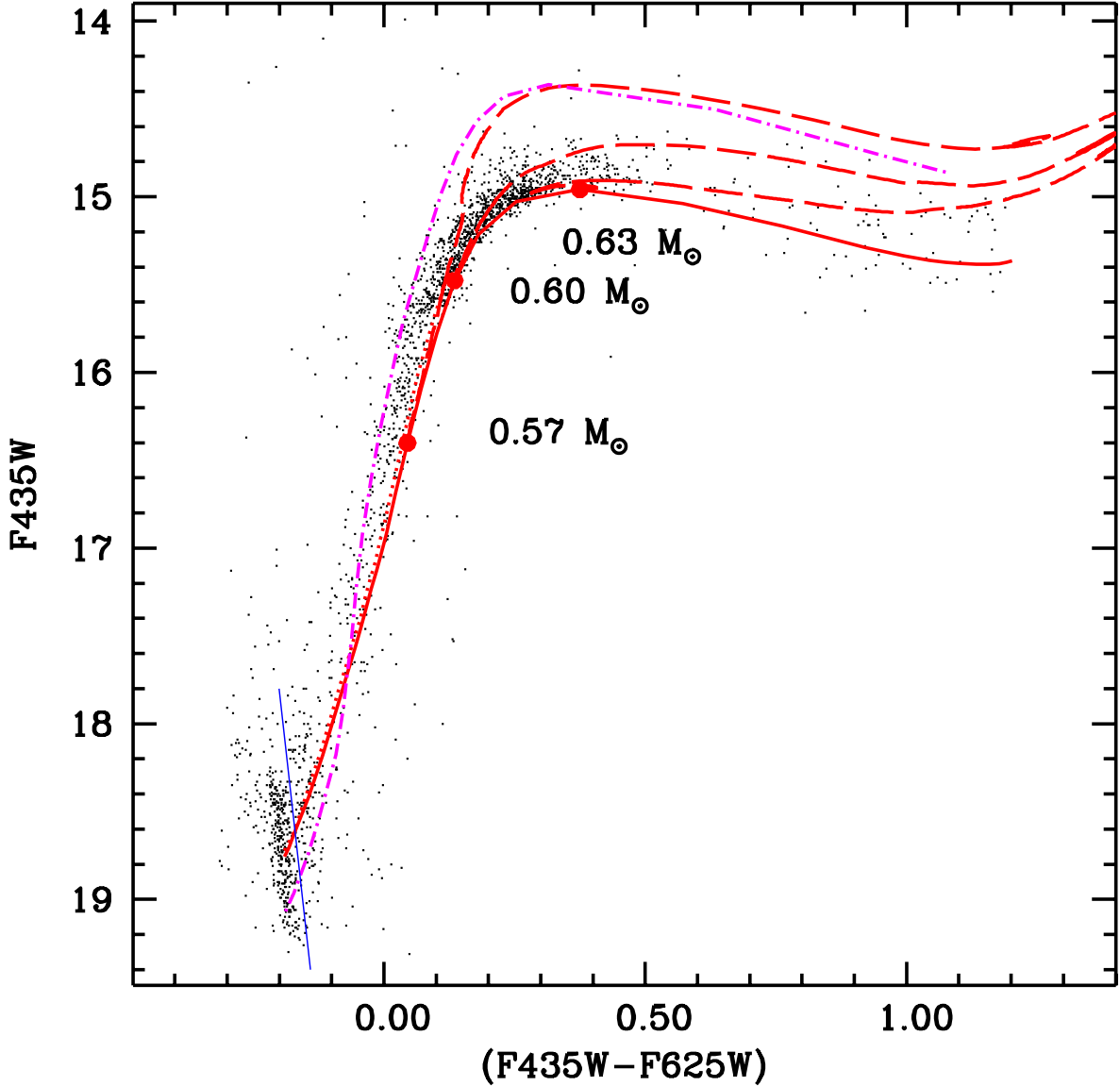}        
\caption{
The HB stars of $\omega$~Cen as selected from the CMD in Fig.~\ref{cmd1}. 
The ZAHB for the reference
population (solid red line) plus three HB evolutionary tracks with the
labelled masses (dashed red lines), and the ZAHB for the He-rich structures (dashed-dotted magenta line)
are also displayed (see text for details). The dotted red line denotes the 
fictitious ZAHB (for $T_{\rm eff}$ larger than 11$\,$000~K) 
that simulated the effect of radiative levitation 
of metals in the atmospheres of hot HB stars (see text for details).  
All ZAHBs and HB tracks
have been corrected for an apparent distance modulus and reddening
equal to $(m-M)_{\rm F435W}$=14.32 and E(F435W$-$F625W)=0.22. 
Filled circles denote the ZAHB location of the three tracks.
The thin solid line in the left-hand corner of the diagram denotes the cool
limit of the \lq{blue clump}\rq\ region.
\label{cmd2}}        
\end{figure}

\begin{figure}        
\plotone{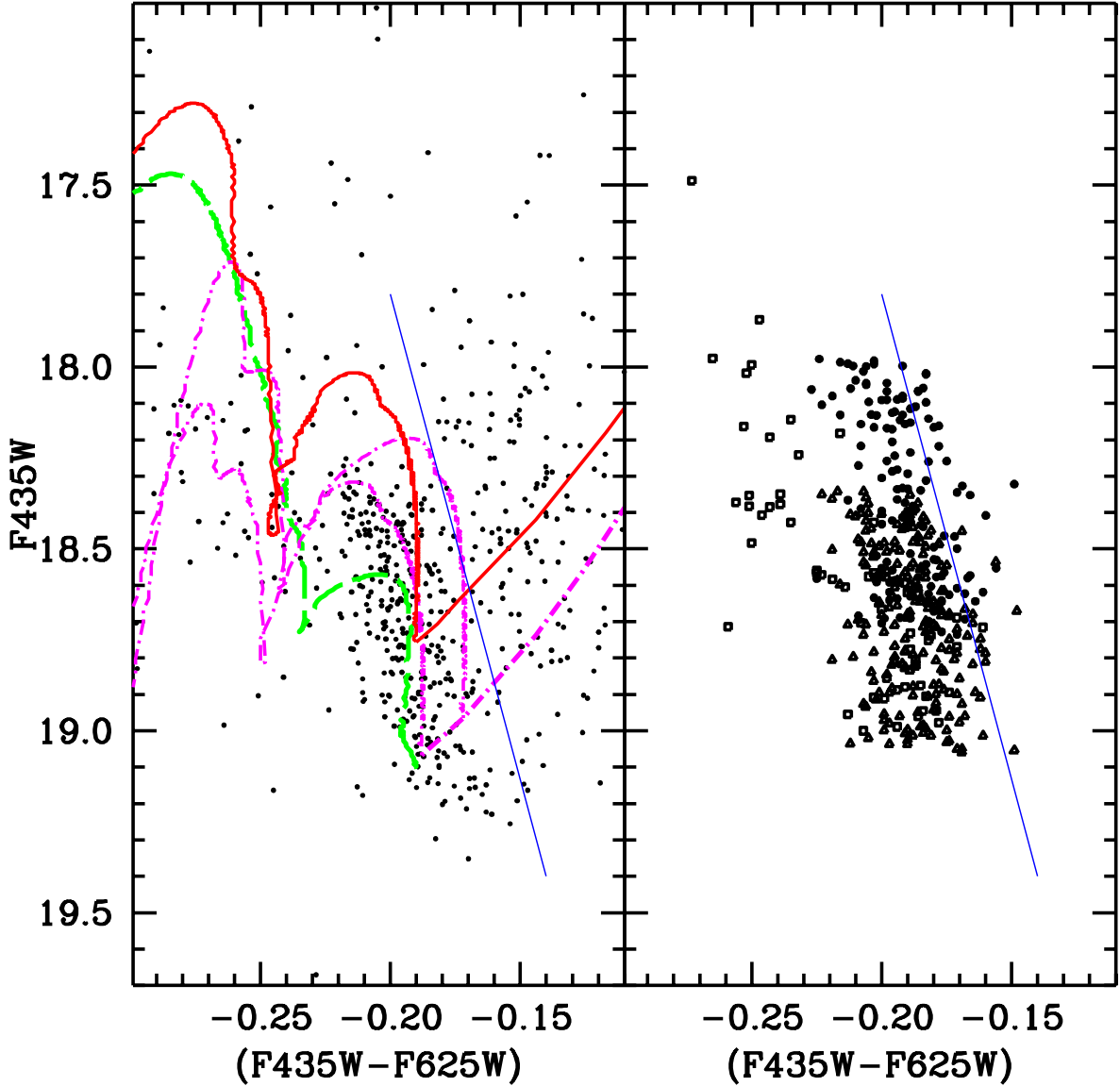}        
\caption{{\sl Left panel}: The \lq{blue clump}\rq\ region in the CMD
of $\omega$~Cen.  The solid red line denotes the ZAHB of the reference
population, the dashed-dotted magenta line represents the He-rich ZAHB,  
and the thin solid blue straight line marks the cool limit of the
\lq{blue clump}\rq\ . The evolutionary track for the hottest reference
HB model with mass equal to $0.491M_\odot$ (solid red line), two He-rich
tracks with masses of $0.460~M_{\odot}$ and $0.461~M_{\odot}$,
respectively (dashed-dotted magenta lines), and an hot flasher model with mass equal
to $0.489~M_{\odot}$ (dashed green line) are also shown. {\sl Right panel}:
The synthetic HB obtained employing the relative numbers of reference
(filled circles), He-rich (open triangles) and hot flasher (open
squares) stars derived by Moehler et al.~(2007) for HB objects with
$T_{\rm eff}$ hotter than 30$\,$000~K (see text for details).
\label{cmd3}}        
\end{figure}

To estimate the relative number of He-rich to reference composition
progeny along the HB, is necessary to multiply the 30\% value obtained
before by the ratio between their HB evolutionary times. From
Fig.~\ref{cmd2} one can notice that along the HB the progeny of the
He-rich population has to be located at F435W fainter than
$\approx$17.2-17.4 mag, given that at brighter magnitudes the He-rich
ZAHB is overluminous compared to the observed HB.  This corresponds to
masses smaller than $\sim0.51 M_{\odot}$. At F435W fainter than
17.4~mag the majority of stars is concentrated in the \lq{blue
clump}\rq, and we therefore assume a typical HB mass equal to
0.460$M_{\odot}$ for the He-rich progeny (see Fig~\ref{cmd3}).

On the other hand, the typical mass of the reference HB component has
to be between $\sim$0.57 and $\sim$0.63~$M_{\odot}$ , corresponding to
the objects along the extremely well populated knee of the observed HB
(see Fig~\ref{cmd2}).  Masses in this range have all extremely similar
central He-burning lifetimes, differing by less than 5\%, and as
representative value we employed the lifetime of the 0.60 $M_{\odot}$
model.

With these estimates the theoretical HB lifetime ratio results to be
equal to $\sim$1.6, the He-rich progeny being longer lived. The
expected ratio between the progeny of the He-rich component and the
progeny of the reference population is therefore equal to $1.6 \times
0.30 = 0.48$.  On the observational side, even if we consider all HB
stars with F435W fainter than 17.4~mag (about 600 objects) being the
progeny of the He-rich population, the observed ratio between He-rich
and reference HB progeny is equal to 0.37$\pm$0.03, lower than the
theoretically expected value.  Any contamination of the HB at 
F435W$>$17.4 by the progeny of the reference population [as
expected in case Moehler et al.~(2007) results were to be valid also
for this cluster field] would exacerbate this discrepancy.

Our result implies that at least $\sim$15\% (about 100 objects) of the
expected He-rich HB stars are unaccounted for. This means that their
RGB progenitors have lost enough mass to avoid central He-ignition,
and ended up along the He-core WD sequence.  There is here an
important connection with Calamida et al.~(2008) result about the WD
counts, that has been obtained by studying exactly the same cluster
fields. These authors conclude that a fraction of He-core WDs ranging
from 80\% (for a mean mass of 0.5 $M_\odot$ for the CO-core and 0.3
$M_\odot$ for the He-core WDs) to 10\% (for a mean mass of 0.5
$M_\odot$ for the CO-core and 0.23 $M_{\odot}$ for the He-core WDs) is
necessary to explain their data.  The 'missing' progeny of the He-rich
MS is a realistic candidate to produce at least a fraction of these
He-core WDs.  The WDs included in Calamida et al.~(2008) calculations
are bright, i.e. they are cooling fast and have been produced by
objects that have just left the HB, or the RGB in case of He-core
WDs. The ratio of the number of 'missing' He-rich HB objects to the
total number of HB stars is therefore a reasonable approximation of the
fraction of bright He-core WDs.  Assuming that all missing He-rich HB stars
have become He-core WDs, they could provide a fraction of the order of
$\sim$5\% of the total WD population.  This contribution would be higher in case
the reference population contributes also to the HB population with
F435W$>$17.4.

We close this section with some considerations about the amount of
mass that RGB stars need to lose to satisfy the constraints posed by
HB models.  The turn off mass of the He-rich population is equal to
$0.59 M_{\odot}$ for an age of 13.5~Gyr.  This means that a RGB mass
loss between only $\sim$ 0.08 and $\sim 0.13~M_{\odot}$ is needed to
populate the HB for F435W$>$17.4.  If the age of the two populations
is decreased to 11~Gyr, the amount of mass lost along the RGB should
be increased by only 0.03 $M_{\odot}$, because of the slightly larger
turn off mass.  In case of the reference population the bulk of the HB
stars should be populating the 'knee' of the observed HB, given the
turn off mass of $0.77 M_{\odot}$ for the assumed cluster age, this
implies a mean mass loss of $\sim$0.17$M_{\odot}$ (that has to be
increased by 0.04~$M_{\odot}$ for an age of 11~Gyr) before settling on
the HB. Very recently McDonald et al.~(2009) have estimated that RGB
stars in $\omega$~Cen lose typically about $0.20-0.25M_\odot$ also
from considerations about the colors of HB stars and assumptions about
the cluster age.

\section{Summary}

Using accurate optical photometric data collected with ACS on board HST, we
have focused on a peculiar clump of stars at the
hottest end of the HB of $\omega$ Cen, that we have termed \lq{blue
clump}\rq. Using theoretical stellar models we have tried to trace the
evolutionary channel(s) able to produce this feature. In particular,
we tried to determine possible links with the peculiar He-rich
sub-population already identified in this cluster.

Our analysis indicates that the \lq{blue clump}\rq\ can be reproduced
by either hot flashers or canonical He-rich HB stars or a mixture of
both types, plus possibly also a contribution, at least for the
brighter part, from the reference HB population.  The objects in the
\lq{blue clump}\rq\ have to span an extremely narrow range of masses,
up to at most only 0.03$M_{\odot}$.  

When we combine the observed population ratios along the MS, the
theoretical evolutionary timescales along MS, RGB and HB, and take
into account the difference in luminosity between the ZAHB of the
reference population and the He-rich one, we find that at least
$\sim$15\% of the expected He-rich HB stars are unaccounted for. The
exact fraction depends on how many stars with F435W$>$17.4 belong to
the reference population. A non-zero contribution implies a larger
fraction of missing He-rich HB stars.  Our results imply that a
fraction of He-rich stars failed to experience the He-flash along the
RGB and have produced instead He-core WD. They should therefore
contribute - at the level of at least about 5\% - to the fraction of
He-core WDs suggested by Calamida et al.\ ~(2008) to explain the
observed ratio between WDs and MS star counts.

\acknowledgments We warmly thank our referee for an accurate reading of the manuscript and for her/his pertinent comments.
 S.C.\ acknowledges the partial financial support of
INAF through the PRIN 2007 grant n. CRA 1.06.10.04: \lq{The local
route to galaxy formation}\rq, and of Ministero della Ricerca
Scientifica e dell'Universit\'a (PRIN-MIUR 2007).  M.S.\ acknowledges
the Astronomy Department of the University of Bologna, and the
Institut D'Estudis Espacials de Catalunya, where large part of this
work was carried out. A.B.\ acknowledges support by the CA.RI.PA.RO.\ foundation, and 
by STScI under  the  2008   graduate  research   assistantship  program.
This research has made use of NASA's
Astrophysics Data System Abstract Service and the SIMBAD database
operated at CDS, Strasbourg, France.



\clearpage

\clearpage

\clearpage


\end{document}